\documentclass[11pt,fleqn]{article} 
\usepackage{psfig}
\title{Gravity-driven  Transport along Cylindrical Topological Defects : \\ Possible Dark Matter and Nearly Frictionless States}  %
\author{Zotin K.-H. Chu}
\date{P.O. Box 39, Distribution Unit, Xihong Road, Urumqi 830000,
China}
\begin{document}           
\maketitle
\doublerulesep=6.5mm        
\baselineskip=6.5mm
\oddsidemargin-1mm         
%
\begin{abstract}
The gravity-driven flow along an annular topological defect
(string) with transversely corrugations is investigated  by using
the verified transition-rate  model and boundary perturbation
method. We found that for certain activation volume and energy
there exists possible frictionless states which might be
associated with the missing momentum of  inertia or dark matter.
\newline

\end{abstract}
%
\bibliographystyle{plain}
\section{Introduction}
%
Quite recently Vitelli {\it et al.} suggested that topological
defects in the cosmic shear can be used as a probe of the
gravitational potential generated by the lensing mass fluctuations
on large scales [1] based on the facts that shear fields due to
weak gravitational lensing have characteristic coherent patterns.
They described the topological defects in shear fields in terms of
the curvature of the surface described by the lensing potential.
In that paper they explored the connection between the theory of
topological defects and the spatial patterns of shear fields due
to weak gravitational lensing. \newline The starting point of
their approach rests on an analogy between gravitational lensing
shear fields, as a probe of structure formation on cosmological
scales, and the anisotropic optical or mechanical response of
materials, as a probe of their inhomogeneous structure on
microscopic scales. As an illustration, the topological defects in
the local shear field of an elastic medium reflect the external
deformations applied to the solid. Similarly for thin liquid
crystal films confined on a curved substrate, the density of
topological defects depends on the inhomogeneous curvature of the
underlying surface [1]. This may also allow them to infer how the
dark matter is concentrated around galaxies and galaxy clusters,
as well as providing a testing ground for dark energy and modified
gravity theories [2,3].
 \newline
Above mentioned or  borrowed analogy is one motivation for our
present study.  The other motivation is related to the possible
dark matter associated with possible superfluidity formation after
shear-thinning. While superflow in a state of matter possessing a
shear modulus might initially seem untenable, experimental claims
for precisely this phenomenon in solid $^4$He now abound [4].
Reported in the experiments of Kim and Chan [5] was a dramatic
change below 200 mK in the period of a torsional oscillator
containing solid $^4$He. Because superfluids come out of
equilibrium and detach from the walls of the rotated container,
they are expected to give rise to a period shift in such a
geometry, assuming, of course, the rotation velocity is less than
the critical velocity to create a vortex. The result is a {\it
missing moment of inertia} (MMI) [4] and hence the period of
oscillation decreases. The magnitude of the MMI is a direct
measure of the superfluid fraction. However, the present author
likes to link this MMI which occurs as there is formation of
superfluidity with the possible formation of dark matter.
\newline
Meanwhile, researchers have been interested in the question of how
matter responds to an external mechanical load. External loads
cause transport, in Newtonian or various types of non-Newtonian
ways. Amorphous matter, composed of polymers, metals, or ceramics,
can deform under mechanical loads, and the nature of the response
to loads often dictates the choice of matter in various
applications. The nature of all of these responses depends on both
the temperature and loading rate.\newline
To the best knowledge of the author, the simplest model that makes
a prediction for the rate and temperature dependence of shear
yielding is the rate-state model of stress-biased thermal
activation [6-8]. Structural rearrangement is associated with a
single energy barrier $E$ that is lowered or raised linearly by an
applied stress $\sigma$ : $R_{\pm}=\nu_0 \exp[-E/(k_B T)] \exp[\pm
\sigma V^*/(k_B T)],$ where $k_B$ is the Boltzmann constant,
$\nu_0$ is an attempt frequency and $V^*$ is a constant called the
'activation volume'. In amorphous  matter, the transition rates are
negligible at zero stress. Thus, at finite stress one needs to
consider only the rate $R_{+}$ of transitions in the direction
aided by stress.\newline The linear dependence will always
correctly describe small changes in the barrier height, since it
is simply the first term in the Taylor expansion of the barrier
height as a function of load. It is thus appropriate when the
barrier height changes only slightly before the system escapes the
local energy minimum. This situation occurs at higher
temperatures; for example, Newtonian transport is obtained in the
rate-state model in the limit where the system experiences only
small changes in the barrier height before thermally escaping the
energy minimum. As the temperature decreases, larger changes in
the barrier height occur before the system escapes the energy
minimum (giving rise to, for example, non-Newtonian transport). In
this regime, the linear dependence is not necessarily appropriate,
and can lead to inaccurate modeling.  To be precise, at low shear
rates ($\dot{\gamma} \le \dot{\gamma}_c$), the system behaves as a
power law shear-thinning material while, at high shear rates, the
stress varies affinely with the shear rate. These two regimes
correspond to two stable branches of stationary states, for which
data obtained by imposing either $\sigma$ or $\dot{\gamma}$
exactly superpose.
\newline
In this short paper,  motivated by the analogy used in [1], we
shall adopt the verified transition-rate-state model [6-8] to
study the gravity-driven transport of cosmic textures (presumed to
be amorphous) within a corrugated annular (cosmic) string. The
possible nearly frictionless states due to strong shear-thinning
will be relevant to the dark matter formation as mentioned above
(considering the MMI [4]). To obtain the law of shear-thinning
matter for explaining the too rapid annealing at the earliest
time, because the relaxation at the beginning was steeper than
could be explained by the bimolecular law, a hyperbolic sine law
between the shear (strain) rate : $\dot{\gamma}$ and  shear stress
: $\tau$ was proposed and the close agreement with experimental
data was obtained. This model has sound physical foundation from
the thermal activation process [6-8] (a kind of (quantum)
tunneling which relates to the matter rearranging by surmounting a
potential energy barrier was discussed therein). With this model
we can associate the (shear-thinning) fluid with the momentum
transfer between neighboring atomic clusters on the microscopic
scale and reveals the atomic interaction in the relaxation of flow
with dissipation (the momentum transfer depends on the activation
(shear) volume : $V^*\equiv V_h$ which is associated with the center
distance between atoms and is equal to $k_B T/\tau_0$ ($T$ is
temperature in Kelvin, and $\tau_0$ a constant with the dimension
of stress).
\newline To consider the more realistic but complicated  boundary
conditions in the walls of the annular (cosmic) string, however,
we will use the boundary perturbation technique [10] to handle the
presumed wavy-roughness along the walls of the annular (cosmic)
string. To obtain the analytical and approximate solutions, here,
the roughness is only introduced in the radial or transverse
direction. The relevant boundary conditions along the wavy-rough
surfaces will be prescribed below.  We shall describe our approach
after this section : Introduction with the focus upon the boundary
perturbation method. The approximate expression of the transport
is then demonstrated at the end. Finally, we will illustrate our
results into two figures and give discussions therein.
\section{Theoretical Formulations} %
We shall consider a steady transport of the (shear-thinning)
amorphous matter in a wavy-rough annular (cosmic) string of $r_1$
(mean-averaged inner radius) with the inner wall being a fixed
wavy-rough surface : $r=r_1+\epsilon \sin(k \theta+\beta)$ and
$r_2$ (mean-averaged outer radius) with the outer wall being a
fixed wavy-rough surface : $r=r_2+\epsilon \sin(k \theta)$, where
$\epsilon$ is the amplitude of the (wavy) roughness, $\beta$ is
the phase shift between two walls, and the roughness wave number :
$k=2\pi /L $ ($L$ is the wavelength of the surface modulation in
transverse direction). 
\newline Firstly, this amorphous matter (composed of  cosmic textures)  can be expressed
as [6-8]
 $\dot{\gamma}=\dot{\gamma}_0  \sinh(\tau/\tau_0)$,
where $\dot{\gamma}$ is the shear rate, $\tau$ is the shear
stress, and $\dot{\gamma}_0 (\equiv C_k  k_B T \exp(-\Delta E/k_B
T)/h$) is with the dimension of the shear rate; here $C_k \equiv 2
V_h/V_m$ is a constant relating rate of strain to the jump
frequency ($V_h=\lambda_2\lambda_3\lambda$,
$V_m=\lambda_2\lambda_3\lambda_1$, $\lambda_2 \lambda_3$ is the
cross-section of the transport unit on which the shear stress
acts, $\lambda$ is the distance  jumped  on each relaxation,
$\lambda_1$ is the perpendicular distance between two neighboring
layers of particles sliding past each other), accounting for the
interchain co-operation required, $h$ is the Planck constant,
$\Delta E$ is the activation energy. In fact, the force balance
gives the shear stress at a radius $r$ as $\tau=-(r \,\delta{\cal
G})/2$ [6-8]. $\delta{\cal G}$ is the net effective gravity
forcing along the transport (or tube-axis : $z$-axis) direction
(considering $dz$ element).\newline Introducing the forcing
parameter
$\Phi = -(r_2/2\tau_0) \delta{\cal G}$
then we have
 $\dot{\gamma}= \dot{\gamma}_0  \sinh ({\Phi r}/{r_2})$.
As $\dot{\gamma}=- du/dr$ ($u$ is the velocity of the transport in
the longitudinal ($z$-)direction of the annular (cosmic) string),
after integration, we obtain
\begin{equation}
 u=u_s +\frac{\dot{\gamma}_0 r_2}{\Phi} [\cosh \Phi - \cosh (\frac{\Phi r}{r_2})],
\end{equation}
here, $u_s (\equiv u_{slip})$ is the velocity over the (inner or
outer) surface of the annular (cosmic) string, which is determined
by the boundary condition. We noticed that  a general boundary
condition for transport over a solid surface [9] was
\begin{equation}
 \delta u=L_s^0 \dot{\gamma}
 (1-\frac{\dot{\gamma}}{\dot{\gamma}_c})^{-1/2},
\end{equation}
where  $\delta u$ is the velocity jump over the solid surface,
$L_s^0$ is a constant slip length, $\dot{\gamma}_c$ is the
critical shear rate at which the slip length diverges. The slip
(velocity) boundary condition above (related to the slip length)
is closely linked to  the mean free path of the particles together
with a  geometry-dependent factor (it is the quantum-mechanical
scattering of Bogoliubov quasiparticles which is responsible for
the loss of transverse momentum transfer to the container walls [10]).
The value of $\dot{\gamma}_c$ is a function of the corrugation of
interfacial energy.  \newline With the slip boundary condition
[9], we can derive the velocity fields and transport rates
along the wavy-rough annular (cosmic) string below using the verified
boundary perturbation technique [11] and dimensionless analysis.
We firstly select $L_s^0$ to be the characteristic length scale
and set $r'=r/L_s^0$, $R_1=r_1/L_s^0$, $R_2=r_2/L_s^0$,
$\epsilon'=\epsilon/L_s^0$. After this, for simplicity, we drop
all the primes. It means, now, $r$, $R_1$, $R_2$ and $\epsilon$
become dimensionless ($\Phi$ and $\dot{\gamma}$ also follow). The
wavy boundaries are prescribed as $r=R_2+\epsilon \sin(k\theta)$
and $r=R_1+\epsilon \sin(k\theta+\beta)$ and the presumed steady
transport is along the $z$-direction (microannulus-axis
direction).
\subsection{Boundary Perturbation}
Along the outer boundary (the same treatment below could also be
applied to
the inner boundary), we have
 $\dot{\gamma}=(d u)/(d n)|_{{\mbox{\small on walls}}}$.
Here, $n$ means the  normal. Let $u$ be expanded in $\epsilon$ :
 $$u= u_0 +\epsilon u_1 + \epsilon^2 u_2 + \cdots,$$
and on the boundary, we expand $u(r_0+\epsilon dr,
\theta(=\theta_0))$ into
\begin{displaymath}
u(r,\theta) |_{(r_0+\epsilon dr,\,\theta_0)} =u(r_0,\theta)+\epsilon
[dr \,u_r (r_0,\theta)]+ \epsilon^2 [\frac{dr^2}{2}
u_{rr}(r_0,\theta)]+\cdots=
\end{displaymath}
\begin{equation}
  \{u_{slip} +\frac{\dot{\gamma} R_2}{\Phi} [\cosh \Phi - \cosh (\frac{\Phi
 r}{R_2})]\}|_{{\mbox{\small on walls}}}, \hspace*{6mm} r_0 \equiv
 R_1, R_2;
\end{equation}
where
\begin{equation}
 u_{slip}|_{{\mbox{\small on walls}}}=L_s^0 \{\dot{\gamma}
 [(1-\frac{\dot{\gamma}}{\dot{\gamma}_c})^{-1/2}]\}
 |_{{\mbox{\small on walls}}}, 
\end{equation}
Now, on the outer wall (cf. [11])
\begin{displaymath}
 \dot{\gamma}=\frac{du}{dn}=\nabla u \cdot \frac{\nabla (r-R_2-\epsilon
\sin(k\theta))}{| \nabla (r-R_2-\epsilon \sin(k\theta)) |}
=[1+\epsilon^2 \frac{k^2}{r^2}  \cos^2 (k\theta)]^{-\frac{1}{2}}
[u_r |_{(R_2+\epsilon dr,\theta)} -
\end{displaymath}
\begin{displaymath}  
 \hspace*{12mm} \epsilon \frac{k}{r^2}
\cos(k\theta) u_{\theta} |_{(R_2+\epsilon dr,\theta)}
]=u_{0_r}|_{R_2} +\epsilon [u_{1_r}|_{R_2} +u_{0_{rr}}|_{R_2}
\sin(k\theta)-
\end{displaymath}
\begin{displaymath}
  \hspace*{12mm}  \frac{k}{r^2} u_{0_{\theta}}|_{R_2} \cos(k\theta)]+\epsilon^2 [-\frac{1}{2} \frac{k^2}{r^2} \cos^2
(k\theta) u_{0_r}|_{R_2} + u_{2_r}|_{R_2} + u_{1_{rr}}|_{R_2} \sin(k\theta)+ 
\end{displaymath}
\begin{equation}
   \hspace*{12mm} \frac{1}{2} u_{0_{rrr}}|_{R_2} \sin^2 (k\theta) -\frac{k}{r^2}
\cos(k\theta) (u_{1_{\theta}}|_{R_2} + u_{0_{\theta r}}|_{R_2}
\sin(k\theta) )] + O(\epsilon^3 ) .
\end{equation}
Considering $L_s^0 \sim R_1,R_2 \gg \epsilon$ case, we also
presume $\sinh\Phi \ll \dot{\gamma}_c/\dot{\gamma_0}$.
With equations (1) and (5), using the definition of
$\dot{\gamma}$, we can derive the velocity field ($u$) up to the
second order :
\begin{displaymath}
u(r,\theta)=-(R_2 \dot{\gamma}_0/\Phi) \{\cosh
(\Phi r/R_2)-\cosh\Phi\, [1+\epsilon^2 \Phi^2 \sin^2
(k\theta)/(2 R_2^2)]+
\end{displaymath}
\begin{displaymath}
 \hspace*{12mm} \epsilon \Phi \sinh \Phi \,
\sin(k\theta)/R_2\}+u_{slip}|_{r=R_2+\epsilon \sin (k\theta)}.
\end{displaymath}
The key point is to firstly obtain the slip velocity along the
boundaries or surfaces.
After lengthy mathematical manipulations, we obtain %
the velocity fields (up to the second order) and then we can
integrate them with respect to the cross-section to get the transport (volume
flow) rate ($Q$, also up to the second order here) :
\begin{displaymath} 
  Q=\int_0^{\theta_p} \int_{R_1+\epsilon \sin(k\theta+\beta)}^{R_2+\epsilon \sin(k\theta)}
 u(r,\theta) r
 dr d\theta =Q_{0} +\epsilon\,Q_{p_0}+\epsilon^2\,Q_{p_2}.
\end{displaymath}
In fact, the approximate (up to the second order) net transport (volume flow)
rate  reads :
\begin{displaymath}
 Q=\pi \dot{\gamma}_0 \{L_s^0 (R_2^2-R_1^2)  \sinh\Phi \,
 (1-\frac{\sinh\Phi}{\dot{\gamma}_c/\dot{\gamma_0}})^{-1/2}+
 \frac{R_2}{\Phi}[(R_2^2-R_1^2)\cosh\Phi-\frac{2}{\Phi}(R_2^2 \sinh \Phi-
\end{displaymath}
\begin{displaymath}
  R_1 R_2 \sinh(\Phi \frac{R_1}{R_2}))+ \frac{2 R_2^2}{\Phi^2}(\cosh\Phi-\cosh(\Phi \frac{R_1}{R_2}))]\}+
  \epsilon^2 \{\frac{\pi}{2} u_{slip_0} (R_2^2-R_1^2)+
\end{displaymath}
\begin{displaymath}
 L_s^0 \frac{\pi}{4} \dot{\gamma}_0  \sinh \Phi (1+\frac{\sinh\Phi}{\dot{\gamma}_c/\dot{\gamma_0}})
 (-k^2+\Phi^2)[1-(\frac{R_1}{R_2})^2]+\frac{\pi}{2}\dot{\gamma}_0
 [R_1 \sinh (\frac{R_1}{R_2} \Phi)-R_2 \sinh \Phi]-
\end{displaymath}
\begin{displaymath}
 \frac{\pi}{2}  \dot{\gamma}_0 \frac{R_2}{\Phi} [ \cosh\Phi - \cosh (\Phi
  \frac{R_1}{R_2})]+ \frac{\pi}{4}  \dot{\gamma}_0 \Phi \cosh\Phi [R_2  - \frac{R_1^2}{R_2}
  ]+
\end{displaymath}
\begin{displaymath}
  \pi \dot{\gamma}_0 \{[\sinh\Phi+L_s^0 \cosh\Phi
  (1+\frac{\sinh\Phi}{\dot{\gamma}_c/\dot{\gamma_0}})] (R_2-R_1 \cos\beta
  )\}+\frac{\pi}{2}\dot{\gamma}_0 \frac{R_2}{\Phi} \cosh \Phi+
\end{displaymath}
\begin{equation}
  L_s^0\frac{\pi}{4} \Phi^2 \dot{\gamma}_0 \frac{\cosh\Phi}{\dot{\gamma}_c/\dot{\gamma}_0}[1
-(\frac{R_1}{R_2})^2
 ]\} \cosh\Phi.
\end{equation}
Here,
\begin{equation}
 u_{{slip}_0}= L_s^0 \dot{\gamma}_0 [\sinh\Phi(1-\frac{\sinh\Phi}{
 \dot{\gamma}_c/\dot{\gamma}_0})^{-1/2}].
\end{equation}
\section{Results and Discussions}
We firstly check the roughness effect (or combination of curvature and confinement effects [12-13])
upon the gravity-driven transport via strongly shearing
because there are no available experimental data and numerical
simulations for the same geometric configuration (annular (cosmic) string with wavy corrugations
in transverse  direction). With a
series of forcings (due to imposed gravity forcings) :
$\Phi\equiv R_2 (\delta {\cal G})/(2\tau_0)$, we can determine the enhanced
shear rates ($d\gamma/dt$) due to gravity forcings. From equation (5), we
have (up to the first order)
\begin{equation}
 \frac{d\gamma}{dt}=\frac{d\gamma_0}{dt} [ \sinh \Phi+\epsilon
 \sin(k\theta) \frac{\Phi}{R_2} \cosh \Phi].
\end{equation}
The parameters
are fixed below (the orientation effect : $\sin(k\theta)$ is fixed
here). $r_2$ (the mean outer radius) is selected as the same as
the slip length $L_s^0$. The amplitude of wavy roughness
can be tuned easily.
The effect of wavy-roughness is significant once the forcing ($\Phi$) is rather
large (the maximum is of the order of magnitude of $\epsilon [\Phi
\tanh(\Phi)/R_2]$). \newline If we select a (fixed) temperature,
 then from the expression of $\tau_0$, we
can obtain the shear stress $\tau$ corresponding to above gravity forcings
($\Phi$) :
\begin{equation}
 \tau =\tau_0 \sinh^{-1} [\sinh(\Phi)+\epsilon
 \sin(k\theta) \frac{\Phi}{R_2} \cosh(\Phi)].
\end{equation}
There is no doubt that the orientation effect ($\theta$) is also
present for the amorphous matter. For illustration below, we only consider the maximum case : $|\sin(k\theta)|=1$.
We shall demonstrate our transport results below. The wave number
of roughness in transverse direction is fixed to be $10$ (presumed
to be the same for both walls of the annular (cosmic) string) here.
\newline
Now, we start to examine the temperature effect. We fix the
forcing $\Phi$ to be $1$ as its effect is of the order $O(1)$ for
the shear rate (cf. Fig. 2). As the gravity forcing ($\delta {\cal G}$) might depend on the temperature
($\delta {\cal G}=2\tau_0 \Phi/R_2$,
$\tau_0\equiv \tau_0 (T)$, $V^*(\equiv V_h)$ is presumed to be temperature
independent here for simplicity).
Note that, according to [7], $V^*=3 V \delta \gamma/2$
for certain matter during an activation event [7], where $V$ is
the deformation volume, $\delta\gamma$ is the increment of shear
strain.
\newline
As the primary interest of present study is related to the
possible phase transition [14-16] or formation of superfluidity
(presumed to be relevant to the formation of dark matter mentioned
in Introduction) due to strong shearing, we shall present our main
results in the following. We performed intensive calculations or
manipulations of related physical and geometric parameters,
considering a hot big-bang universe [14] and examine what happens
as it expands and cools through the transition temperature $T_c$.
The selected temperature range and the activation anergy follows
this reasoning.  Note that in unified models of weak and
electromagnetic interactions $T_c$ is of the order of the square
root of the Fermi coupling constant [14], $G_F^{1/2}$, i.e. a few
hundred GeV. Thus the transition occurs when the universe is aged
between $l0^{-10}$ and $10^{-12}$ seconds and far above nuclear
densities [14,17]. One possible superfluidity formation regime is
demonstrated in Fig. 1.
The activation energy ($\Delta E$) is $10^{-10}$ Joule ($\sim 1$ GeV) and the activation volume is $10^{-9}$ m$^3$. In
fact, all the results shown in this figure depend on
$\dot{\gamma}_0$ and are thus very sensitive to $\Delta E$.
Here $C_k=2$ and the sudden jump of the shear stress (directly
linked to the friction) occurring around $T\sim 10^{11}$
$^{\circ}$K  could be the transition temperature for the selected
$\Delta E$ and $C_k$. There is a sudden friction drop around  two
orders of magnitude below $T\sim 10^{11}$ $^{\circ}$K and it is
almost frictionless below $T\sim 10^{10}$ $^{\circ}$K. If we
borrow the analogy from the MMI [4] (the rotational properties of
a superfluid as well as a supersolid in which some of the
particles remain still while the rest of them rotate with the
container) we can  identify the formation of superfluidity as the
possible dark matter formation as the mass is missing within this
regime.\newline The possible reasoning for this formation can be
illustrated in Fig. 2. It could be due to the strong shearing
driven by larger gravity forcings along a confined (cosmic) string
or topological defect. The shear-thinning (the viscosity
diminishes with increasing shear rate) reduces the viscosity
significantly. One possible outcome for almost vanishing viscosity
is the nearly frictionless transport. It  seems to us that the
formation of dark matter is a genuinely  dynamic effect. As to the
question whether the internal structure of defects lead to
persistent currents in their cores [17] is yet open. We shall
investigate the evolution of the (cosmic) string [14,17] in the
future.

\newpage

\psfig{file=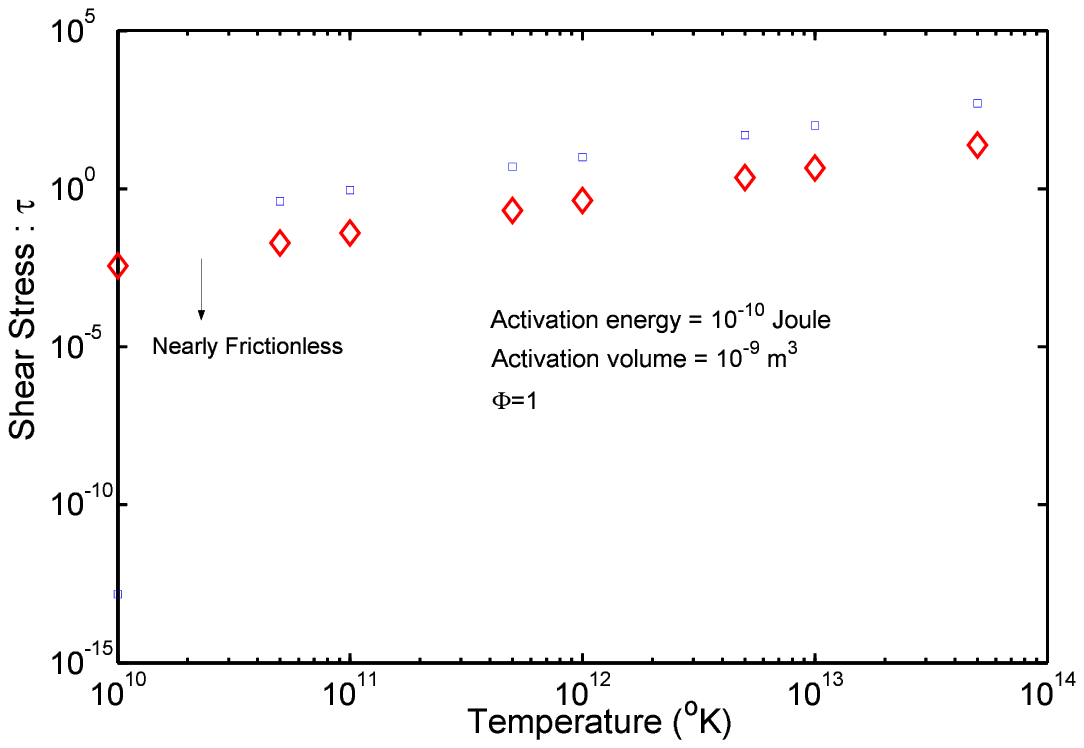,bbllx=-1.5cm,bblly=18cm,bburx=12cm,bbury=26cm,rheight=8cm,rwidth=12cm,clip=}

\begin{figure} [h]
\hspace*{7mm} Fig. 1. Comparison of calculated (shear) stresses
using an activation energy
 $10^{-10}$ J or \newline \hspace*{8mm} $\sim 1$ GeV. There is a sharp
decrease of shear stress around T $\sim 10^{11}$ K.
 Below $10^{10}$ K, \newline \hspace*{8mm} the transport of amorphous matter is nearly
frictionless.
\end{figure}

\newpage

\psfig{file=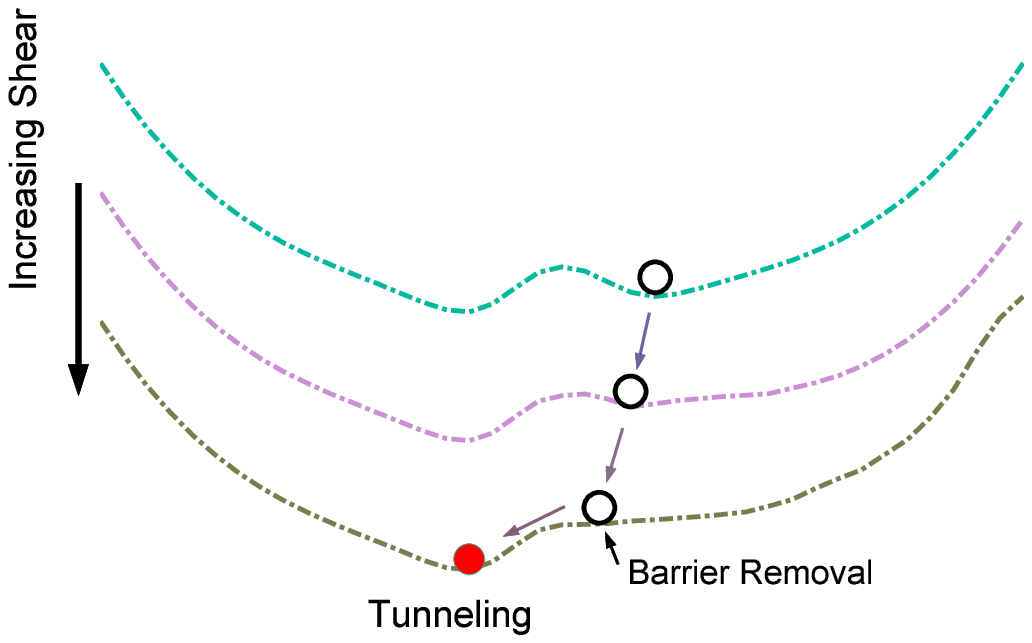,bbllx=-1.0cm,bblly=18cm,bburx=12cm,bbury=26cm,rheight=8cm,rwidth=10cm,clip=}

\begin{figure} [h]
\hspace*{8mm} Fig. 2. Increasing shear causes a local energy
minimum to flatten until it disappears
\newline  \hspace*{9mm} (energy barrier removal or quantum-like
tunneling). The structural contribution \newline  \hspace*{9mm}
to the shear stress is referred to shear-thinning.
\end{figure}

\end{document}